**Paweł Max Maksym (1983-2013) Polish Astronomer and Film-maker**

Costantino Sigismondi, ICRANet, Pescara and Observatorio Nacional, Rio de Janeiro

sigismondi@icra.it

**Abstract:**

The sudden and untimely death of Paweł Max Maksym will not diminish his contributions to the field of occultation astronomy, and to Polish society in general. Founder of the Pope Silvester II Observatory in Bukowiec, he was also writing a book to introduce children to astronomy. Graduating in Geography with an experimental thesis in Lunar Occultations at Łódź University, Paweł earned a diploma from the prestigious National Film School in Łódź. An expert observer, he mastered the technique of stellar occultations, fostering the scientific activity of the Observatory. A review of his publications in Minor Planet Circulars and on YouTube is presented here.

**Introduction**

It is always sad to write an obituary, as well as very demanding, because a whole life has to be summarised in a few lines with the obvious risk of missing the more important points. It is sadder when the person was a young friend, and even more so because Paweł Maksym died at the age of 29, following a surgical operation, leaving his wife Katarzyna with their son Karol, only two years old. The life of an astronomer should be as long as possible, to allow the person to experience the movements of all the celestial spheres. The astronomical cycle with which he becomes familiar is the Saros which governs eclipses,[1] connected with the 18.6 years which rule lunar occultations.[2] The first Saros is necessary to be acquainted with what happens with it, the second one for starting to observe, the third one to analyse the data and coordinate other observations and so on… and Paweł did not have the opportunity even to complete his second Saros. For Paweł Maksym, the seventh sphere, the one of Saturn, just completed its first orbit when he passed away at the University Hospital of Prof. Barlicki in Łódź. His dates are: born in Łódź on 27th May 1983 and passed away on 13th February 2013.[3] Because of legal procedures his funeral had been delayed until the 22nd February in the Church of Saints Peter and Paul in

---

[1] 18 years 11 days, known already to Chaldean astronomy and reported by Ptolemy in the Almagest. It is the least common multiple of three periods: the synodic, the draconic, and the anomalistic months, F. Chalub, Revista Brasileira de Ensíno de Física, **31**, 1303 (2009). www.sbfisica.org.br/rbef/pdf/311303.pdf

[2] This is the time for a complete nodes' revolution, or draconic period.

[3] http://www.youtube.com/watch?v=JEs4LRH7tAo (TVP Łódź, news of February 13 at 22h)

Łódź, and he was buried in Saint Ann Cemetery. The measure of a life, after all, is not its duration, but its donation.[4]

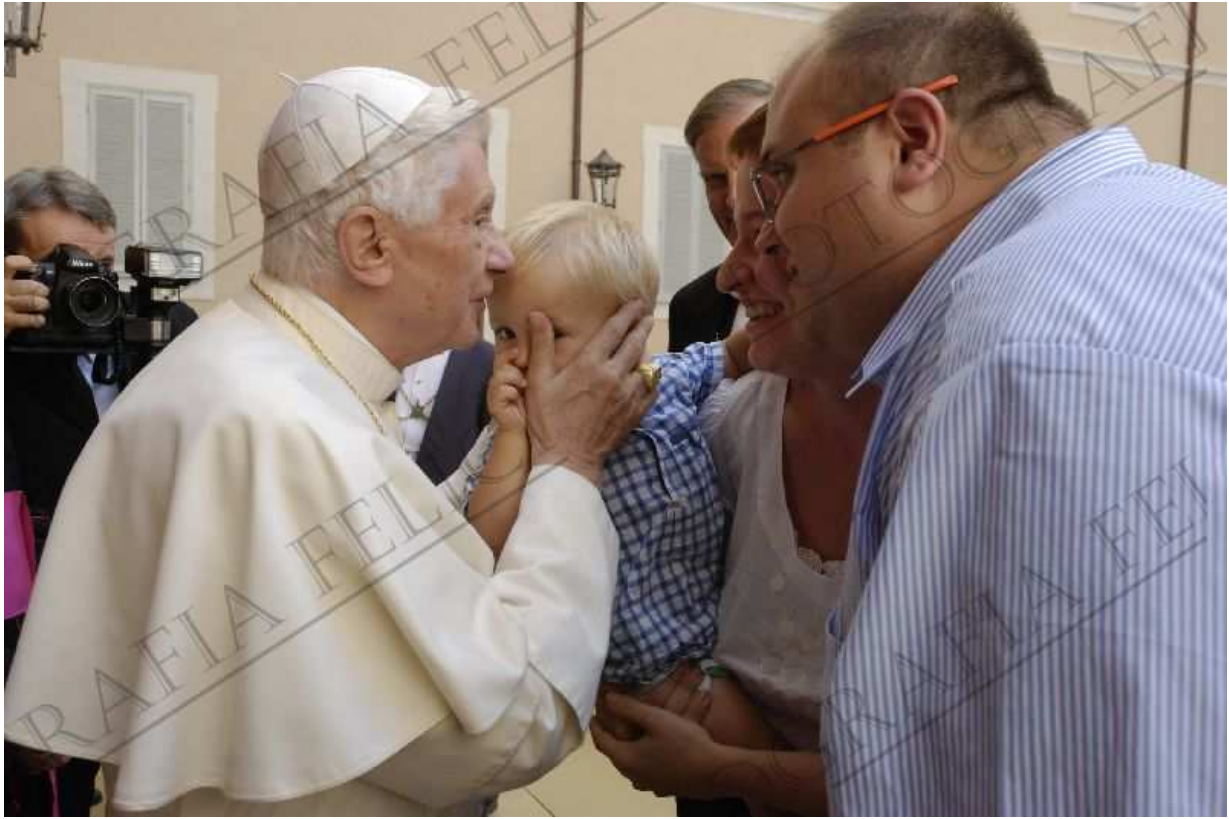

Fig. 1 Paweł Maksym, his wife Katarzyna and their son Karol Josef, receiving a Papal Blessing during their audience with Pope Benedict XVI in Castelgandolfo on August 22, 2012.

**Astronomical work: the world of occultations**

European amateur and professional astronomers engaged in occultation studies are coordinated by IOTA/ES (International Occultation Timing Association/European Section). In Poland there is the Department of Position and Occultations of PAAA, the Polish Amateur Astronomers Association, with which Paweł was also closely associated.[5]

The main objectives of his occultation studies can basically be summarised in a few points:-

---

[4] Corrie Ten Boom (1982-1983)

[5] Paweł brought to my attention that the young Karol Józef Wojtyla, later Pope John Paul II, was inducted into this association in 1938-1939 in his nineteenth year, because in the year 2000, on the day of the Jubilee of Scientists on 25 May, a telescope was donated to the Pope, after he privately expressed this desire, and I was physically in charge of bringing a Schmidt-Cassegrain 8" telescope to the Pope in the Basilica of St. Peter on behalf of the scientific community.

- Lunar grazing occultations: definition of the Cassini zones in the Watts profile

- Lunar total occultations: calibration of the Kaguya profile

- Asteroidal occultations: determination of the size and position of asteroids

- Trans-Neptunian Objects occultations: determination of their position, size and atmospheric pressure.[6]

The first two points were already developed in his Master's degree thesis on Lunar Occultations, already written before the year 2009, and finally presented to the University of Łódź in the month of September 2012.

'In the field' observational campaigns are required in order to go to the exact locations where the star's path relative to Moon intercepts the mountains and valleys of the lunar limb. The stars appear to blink during these passages, giving a precise measurement of the angular amplitude of these valleys. An accuracy comparable with Kaguya laser-altimeter measurements can be locally achieved with these timing data.

The second point was in the observational projects included in the research programme of the Observatory "Pope Silvester II" of Bukowiec, which was inaugurated on the 21st of May 2010, five months after the Kaguya data were made public.[7]

For the third point, the probability of a fixed observatory lying on the path of an asteroidal occultation increases as the limiting magnitude increases. Otherwise it is more common to organize trips with movable instruments to be located on the predicted path, as in the case of lunar grazes.

Paweł equipped the telescopes with increasingly better instruments, starting with a Sony HC96E[8], a PC170 and a VX2100 camera,[9] progressing to a Watec 902H camera.[10] These upgrades corresponded to an improvement in the limiting magnitude for video observations at 25 frames per second up to Mv≈12, using a 20 cm Newtonian telescope.

---

[6] Elliot, J. L. and C. B. Olkin, Probing Planetary Atmospheres with Stellar Occultations, Annual Review of Earth and Planetary Sciences, 24, 89 (1996).

[7] Since November 2, 2009.

[8] http://www.youtube.com/watch?v=Gutyrl6GDN4&list=UUAhmVcvUFXjQoGgliyo2G9g&index=9

[9] http://www.youtube.com/watch?v=m1Al4bZUrHw&list=UUAhmVcvUFXjQoGgliyo2G9g&index=10

[10] http://www.youtube.com/watch?v=yCxDoEj9Ktw&list=UUAhmVcvUFXjQoGgliyo2G9g

The timing resolution is important for lunar occultations since the angular velocity of the Moon is about 0.5 arcsec/s, then a resolution of 1/25 s in the occultation timing corresponds to 0.02 arcsec.

In the case of asteroidal occultations the relative angular velocities of the objects are lower, but also the angular diameters of them are much smaller, so the timing resolution required is always the higher the better. The largest trans-Neptunian objects - at distances ten times larger than the asteroids - offer longer occultations, where the video capability in timing resolution can be reduced in order to reach higher limiting magnitudes.

At the moment the Observatory, equipped with a 25 cm f/6.3 Schmidt Cassegrain telescope was capable of $M_v \approx 12.5$ with 25 fps (0.04 s of maximum integration time), which corresponds to $M_v \approx 17.5$ with a frame rate of one image every 4 seconds.

About the fourth point, the TNO occultations, it is interesting to note also that the light curve of a TNO occultation can give information on the presence and the density of an atmosphere around the TNO. A pressure of a few nano-bar is sufficient to bend the stellar rays by refraction, and some light arrives even during the totality at the shadow plane located tens of Astronomical Units away on the Earth.

The following figure summarises this point:-

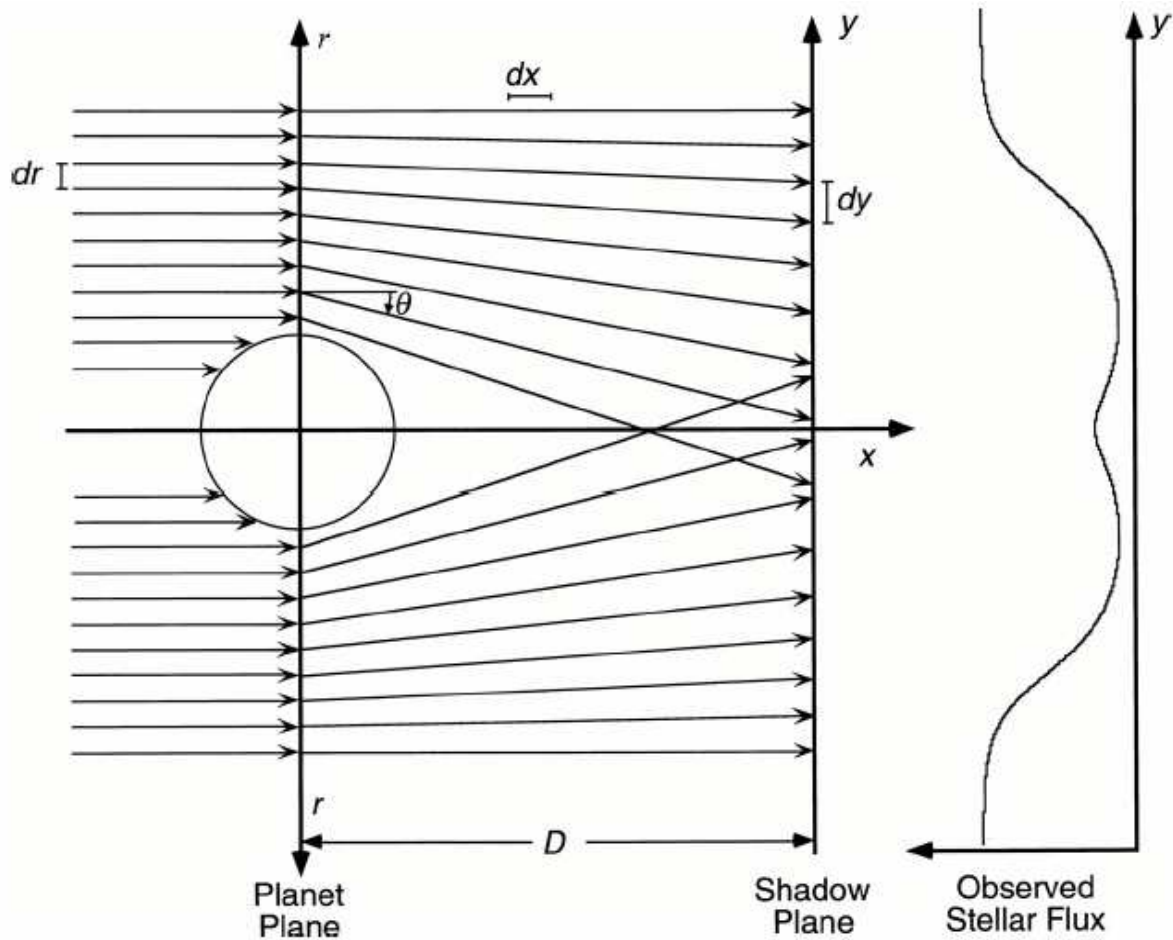

Fig. 2 Refraction of starlight by a planetary atmosphere. Starlight incident from the left encounters a planetary atmosphere and is refracted toward the density gradient as illustrated. The exponential gradient causes the rays to spread, which is seen as a dimming of the star by a distant observer located in the shadow plane. In general, light from both the near and far limb contribute to the light curve (depicted at the far right), although the near-limb contribution is dominant.[11]

**Varuna occultation of January 8, 2013: focus on dwarf planets of our solar system**

As is well known, the technique of stellar occultations is related to the possibility to make observations with portable instrumentation or to wait until the occasion when the fixed observatory is under the eclipse path. The ten biggest Trans Neptunian Objects (TNOs), known up to now, occult several stars brighter than magnitude R = 18.0 every year. The number of occultations ranges from 1-2 in the case of Eris to 100-400 for Ixion;[12] this is a new opportunity especially for amateur astronomers to contribute to the knowledge of these dwarf planets.

---

[11] The figure and the caption are reproduced from of Elliot, J. L. and C. B. Olkin, Probing Planetary Atmospheres with Stellar Occultations, Annual Review of Earth and Planetary Sciences, **24**, 89 (1996).

Using the telescopes at the Pope Silvester II Observatory, Paweł Maksym attempted to observe the Varuna occultation,[13] which was predicted for January 8, 2013. His technical and theoretical preparation, his observational skills, and the equipment that he provided to the Observatory he founded and previously tested with many other occultations, were capable of this very demanding task.

This occultation was first predicted for 8th Jan 2013 by the Rio Team and Bruno Sicardy as a polar region event. In the figure the dots identify each 1000 km along the path of the occultation, or 38.62 s. <> offsets (mas) -45.0 5.0

At 20:25:26.0 RA 07h49m36.9977s DEC+26°25'51".922. C/A =0.179 P/A=14.03 velocity=-25.89 km/s Delta=42.67 the magnitudes R*=16.4 K*=14.6.

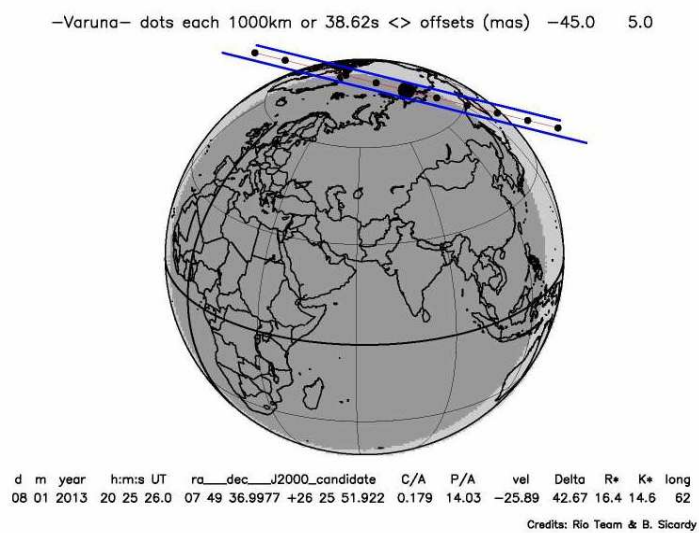

Fig. 3 First prediction of the Varuna occultation of Jan. 8 2013

---

[12] M. Assafin, et al., Astronomy and Astrophysics **541**, A142 (2012), table 9.

[13] In his last message to the PLANOCCULT mailing list he wrote on Jan 9, 2013 at 0:38 TMEC.
Dear "Varunators" ;)
It's unusual for me to report clouds but in case of essential events it could be important. This time I have to do that and report that in Pope Silvester II's Astronomical Observatory in Bukowiec (http://www.oabukowiec.pl/) we were ready for Varuna registration but sky was fully covered by clouds and a typical snowfall also occurred... So, next time I hope. Now it appears that 2013 will be overcasted (mostly when good event occur)...as 2012 was for us.
Starry sky for all of You!
Paweł Maksym

Later, with improved astrometry, a new prediction favourable for central Europe was announced,[14] and this prediction was finally corrected to a zone covering only Japan among the populated countries, just 3 days before.

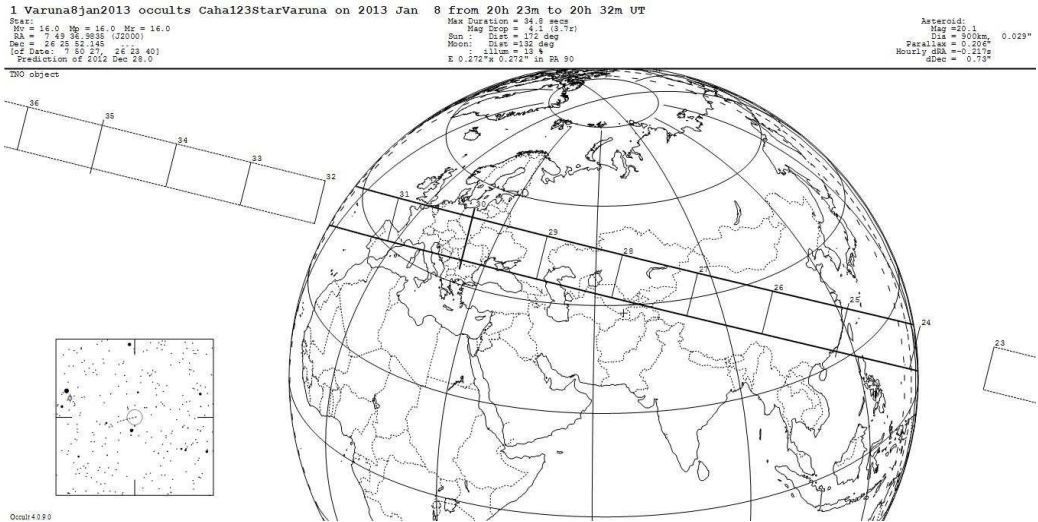

Fig. 4 Second prediction of the Varuna occultation of Jan. 8 2013

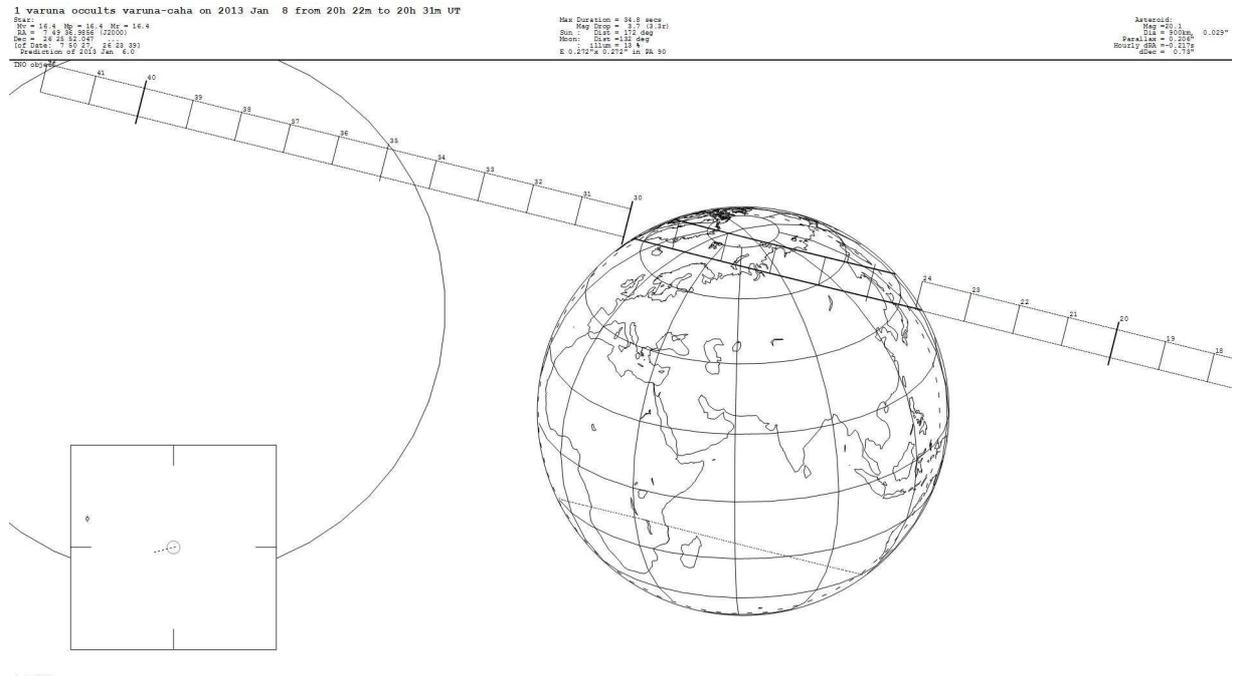

Fig. 5 Last prediction of the Varuna occultation of Jan. 8 2013

---

[14] J. Ortiz, http://www.iaa.es/~ortiz/Varuna20130108.html

Wolfgang Beisker of IOTA/ES was following the development of the predictions to update the European observers.[15] It is also interesting to note that the timing predictions made by Julio Camargo of the Rio Team - as the new astrometric data from the Pic du Midi were made available - showed a shift of 1 m 35s.

The news of two positive (successful) observations of the Varuna occultation from Hiroshima in Japan confirmed the latest predictions to within a few seconds.

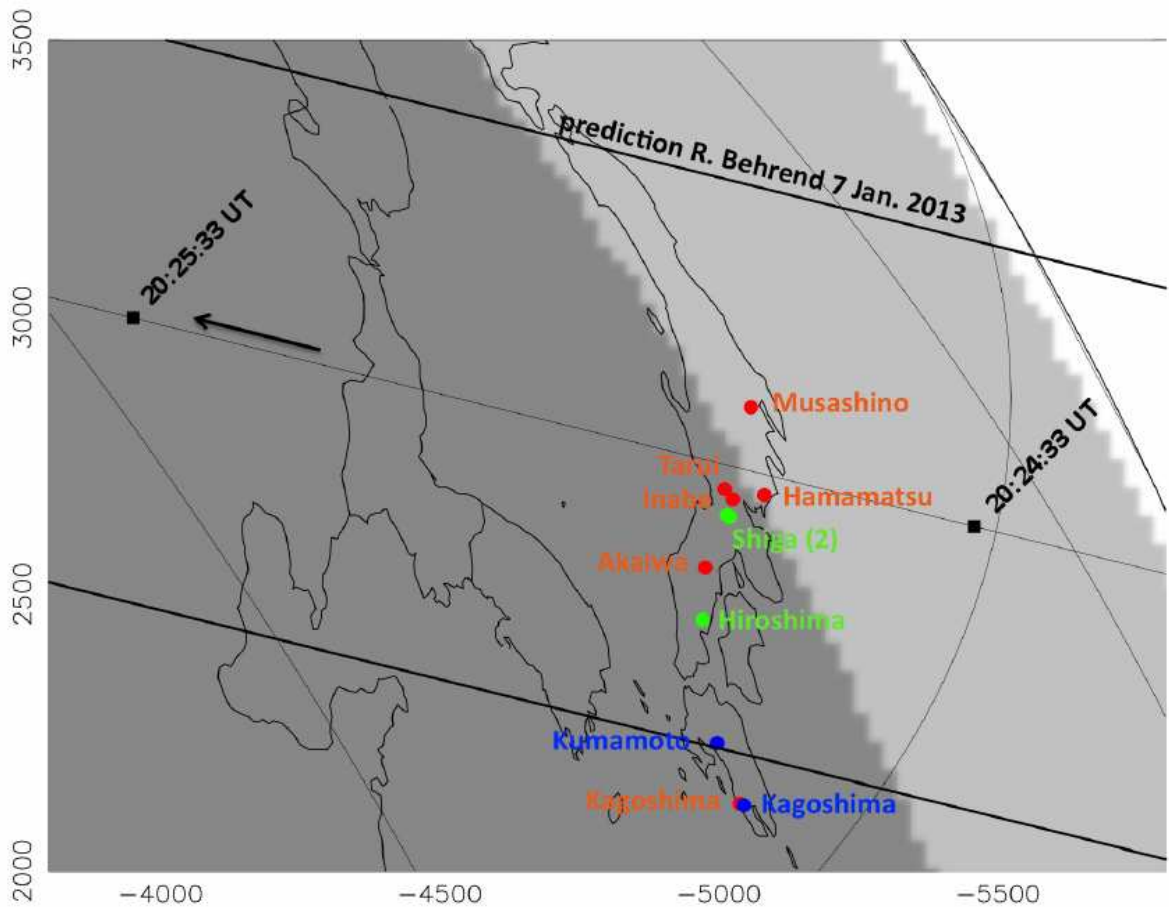

Fig. 6 Observations from Japan of the Varuna occultation of Jan. 8 2013.

---

[15] http://www.iota-es.de/varuna08_01_2013/varuna_08012013.html

The green points are where the occultation was positive, blue points where it was negative and red where it was not possible to obtain data.[16]

**Asteroidal occultations**

As in the case of TNOs, the asteroidal occultations are observed either from fixed observatories, offering larger aperture telescopes with fixed and stable mounts and accurate pointing, or in the field with portable instrumentation.[17]

Paweł Maksym observed several asteroidal occultations, four of them in the first years of activity of the Pope Silvester II Observatory,[18] and one of them resulted in a positive observation.

The umbral path of the naked-eye occultation of delta Ophiuchi by the asteroid (472) Roma was several hundreds of kilometres from the Observatory, but for the importance of the event, like the majority of IOTA/ES observers, he participated in the observational campaign, siting his instruments in the field.

This particular occultation had been followed by many observers along its umbral path,[19] but the ephemerides published in the last few days contained an uncertainty of several kilometres and many observers located near the predicted centreline did not see the occultation.

Paweł Maksym's observations of these asteroidal occultations were included in five Minor Planet Circulars under the MPC Observatory code 244 "Geocentric Occultation Observation".[20]

---

[16] F. Braga-Ribas, Explorando os objectos transnetunianos pelo método de ocultações estelares: predição, observação, Quaoar e os primeiros resultados, Tese de doutorado em Astronomia, ON/Observatoire de Paris (2013), fig. 5.13 (courtesy of Bruno Sicardy).

[17] S. Degenhardt, High Resolution Asteroid Profile by Multi Chord Occultation Observations, http://scottysmightymini.com/PR/HighResAstProfile.pdf

[18] The last communications of asteroidal occultations that Paweł Maksym sent to the PLANOCCULT mailing list, of the European coordinators of occultations projects since the International Year of Astronomy 2009 are listed below, the S2O between parenthesis means that the observation was made at the Pope Silvester II Observatory:

1. DATE: 2009 September 08 STAR: HIP 25965 ASTEROID: 10247 Amphiaraos (Łódź -Mimozy)
2. DATE: 2010 July 08 STAR: HIP 79593 ASTEROID: 472 Roma (Drensteinfurt, Germany)
3. DATE: 2011 Mar 2 STAR: 3UC225-098375 ASTEROID: 4234 Evtushenko (S2O)
4. DATE: 2011 Mar 8 STAR: TYC 1879-00114-1 ASTEROID: 554 Peraga (S2O)
5. DATE: 2011 June 7 STAR: HIP 48340 ASTEROID: 173 Ino (S2O) positive occultation

[19] On YouTube is available among others the video of R. Schoenfeld http://www.youtube.com/watch?v=GZHjJnjYfNQ

[20] Probably for this reason the Silvester II Observatory has not yet been included in the MPC list of observatories with an individual code.

The positive and negative occultations were calculated, reduced to the geocentre and published by D. Herald, G. Blow, D. Dunham, R. Dusser, E. Frappa, T. Hayamizu, J. Manek, M. Soma, J. Talbot, G. Taylor, B. Timerson.[21]

The total number of occultations observed by Paweł Maksym and recorded on the Euraster[22] website is 23, of which 3 were positive and 2 were observed outside Poland: the occultation of delta Ophiuchi by (472) Roma in Germany and the one of (345) Tercidina (in Italy, near Sanremo). The following table[23] summarises all the observervations.

---

[21] http://www.minorplanetcenter.net/iau/ECS/MPCArchive/2007/MPC_20070926.pdf

http://www.minorplanetcenter.net/iau/ECS/MPCArchive/2008/MPC_20080122.pdf

http://www.minorplanetcenter.net/iau/ECS/MPCArchive/2009/MPC_20091231.pdf

http://www.minorplanetcenter.net/iau/ECS/MPCArchive/2011/MPC_20110715.pdf

http://www.minorplanetcenter.net/iau/ECS/MPCArchive/2011/MPC_20110912.pdf

[22] http://www.euraster.net/results/2010/index.html#0708-472

[23] Taken from the website http://www.sky-lab.net/?Occultation_Reports_DB

| OBS | OCC | Date | Star | Asteroid/Planet | SUC | Meth. | Instr. | CC | Observer | UT1 | UT4 | Dur. [s] |
|---|---|---|---|---|---|---|---|---|---|---|---|---|
| 8214 | 2714 | 2011-06-07 | HIP 48340 | (173) Ino | O+ | VID | M250 | PL | P. Maksym et al | 21:00:00 | 21:10:00 | 4.96 |
| 8385 | 2796 | 2011-03-08 | TYC 1879-00114-1 | (554) Peraga | O- | VID | M250 | PL | P. Maksym et al | 20:57:00 | 21:08:00 | |
| 8406 | 2800 | 2011-03-02 | 3UC225-098375 | (4234) Evtushenko | O- | VID | M250 | PL | Pawel Maksym | 22:00:00 | 22:15:00 | |
| 7471 | 2401 | 2010-07-08 | HIP 79593 | (472) Roma | O- | VID | L100 | DE | Pawel Maksym | 21:50:00 | 22:06:00 | |
| 6610 | 2084 | 2009-09-08 | HIP 25965 | (10247) Amphiaraos | O- | VID | M250 | PL | Pawel Maksym | 00:50:00 | 01:05:00 | |
| 6627 | 2089 | 2009-08-24 | TYC 2934-00106-1 | (71) Niobe | O+ | VID | M250 | PL | P. Maksym et al | 23:50:00 | 01:10:00 | 0.48 |
| 5947 | 1837 | 2008-09-04 | TYC 0727-01424-1 | (1144) Oda | O- | VID | M200 | PL | Pawel Maksym | 01:23:00 | 01:31:00 | |
| 5165 | 1541 | 2007-09-21 | 2UCAC 38215341 | (663) Gerlinde | O+ | VID | M200 | PL | Pawel Maksym | 01:42:00 | 01:47:00 | 4.28 |
| 5708 | 1726 | 2007-01-14 | HIP 29196 | (840) Zenobia | O- | VIS | M200 | PL | Pawel Maksym | 16:20:00 | 16:33:00 | |
| 4853 | 1377 | 2006-04-21 | TYC 0406-02150-1 | (15457) 1998 YN6 | O- | VIS | M200 | PL | Pawel Maksym | 00:50:00 | 01:03:00 | |
| 3999 | 1012 | 2005-07-28 | 2UCAC 26766082 | (2397) Lappajarvi | O- | VIS | M200 | PL | Pawel Maksym | 00:23:00 | 00:27:00 | |
| 4206 | 1081 | 2005-03-19 | HIP 43206 | (-) 1999 CO153 | O- | VIS | L70 | PL | Pawel Maksym | 22:00:00 | 23:10:00 | |
| 4279 | 1110 | 2005-02-08 | TYC 0207-00824-1 | (1936) Lugano | O- | VIS | M200 | PL | Pawel Maksym | 23:12:00 | 23:18:30 | |
| 4300 | 1118 | 2005-02-06 | TYC 1344-02003-1 | (1306) Scythia | O- | VIS | M200 | PL | Pawel Maksym | 02:48:00 | 02:52:30 | |
| 3358 | 791 | 2004-09-06 | HIP 84012 | (287) Nephthys | O- | VIS | M200 | PL | Pawel Maksym | 15:21:00 | 15:27:00 | |
| 3394 | 805 | 2004-07-31 | TYC 1688-01854-1 | (849) Ara | O- | VIS | M200 | PL | Pawel Maksym | 00:40:00 | 00:50:00 | |
| 2879 | 652 | 2003-08-26 | TYC 5757-00353-1 | (420) Bertholda | O- | VIS | M110 | PL | Pawel Maksym | 21:30:00 | 22:00:00 | |
| 2969 | 676 | 2003-04-17 | TYC 1368-01752-1 | (407) Arachne | O- | VIS | M120 | PL | Pawel Maksym | 21:14:00 | 21:25:00 | |
| 2234 | 544 | 2002-09-17 | HIP 19388 | (345) Tercidina | O- | VIS | M90 | IT | Pawel Maksym | 00:41:00 | 01:01:00 | |
| 2294 | 547 | 2002-08-29 | HIP 26351 | (1567) Alikoski | O- | VIS | M150 | PL | Pawel Maksym | 23:45:00 | 00:08:00 | |
| 2342 | 565 | 2002-05-12 | TYC 6747-01271-1 | (280) Philia | O- | VIS | M150 | PL | Pawel Maksym | 22:27:00 | 22:48:00 | |
| 2436 | 585 | 2002-02-15 | TYC 2847-00852-1 | (36) Atalante | O- | VIS | M150 | PL | Pawel Maksym | 17:50:00 | 18:15:00 | |
| 2477 | 590 | 2002-02-02 | TYC 4742-00609-1 | (1051) Merope | O- | VIS | M170 | PL | Pawel Maksym | 20:24:00 | 20:49:00 | |

Based on data from: Frappa, E. - European Asteroidal Occultation Results - www.euraster.net

It is to be noted that despite 15 negative observations in a row, Paweł Maksym continued improving his skills and contributing his observations. This shows his awareness of the importance of these observations even if they were negative. After he started to video record occultations he obtained his first positive observation.

A video of the asteroid (71) Niobe occulting a 9.4 magnitude star was the first positive asteroidal occultation uploaded by Paweł Maksym onto YouTube.[24] It was obtained in Bukowiec, 1.79 km from the site of the Pope Silvester II Observatory, at that time in construction. This brief occultation lasted 0.48s. The other positive occultation available on Paweł's videos on YouTube is that of HIP 48340 of magnitude 8.2, occulted by (173) Ino. It was observed from the Pope Silvester II Observatory and its duration of 4.96s was the longest occultation observed by him.[25]

The asteroid (315166) 2007 GA4 discovered on April 6, 2007 by the team of Barbara Dłużewska and the students at the Tadeusz Czacki High School in Warsaw, Poland within the IASC

---

[24] http://www.youtube.com/watch?v=wh7lerr0pMM&list=UUAhmVcvUFXjQoGgliyo2G9g&index=6

[25] http://www.youtube.com/watch?v=wh7lerr0pMM&list=UUAhmVcvUFXjQoGgliyo2G9g&index=5

project,[26] has been proposed through the official channels to be renamed (315166) Paweł Maksym.[27]

**Classical lunar grazes**

The observation of lunar grazes is the classical activity for all occultation observers. The reference for this kind of observation is the book by H. Povenmire, where several fun experiences are included.[28] Paweł published two grazing occultations on YouTube.

The first graze he put on YouTube was that of SAO77818 of April 21, 2007 mv=6.7,[29] the second was the graze of the star ZC1298 of Mv=6.4 on April 13, 2008.[30] In both observations the participants of the in the field mission are presented. Interviews were conducted before and after each observation, making the event an internet performance, with real scientific value.

After the publication of the Kaguya lunar profile in November 2009[31] the interest in the study of the lunar Cassini regions has reduced, but lunar grazes remain a very good tool for calibrating the Kaguya profile and for discovering very close double stars, and eventually measuring stellar diameters.

---

[26] J. Patrick Miller, et al., Astronomy Education Review, **7**, 57-83 (2008). An International Asteroid Search Campaign Internet-Based Hands-On Research Program for High Schools and Colleges, in Collaboration with the Hands-On Universe Project. http://aer.aas.org/resource/1/aerscz/v7/i1/p57_s1

[27] From: "Miller, Patrick" To: Barbara Dłużewska CC: Center for Theoretical Physics Subject: RE: Asteroid Discovery -- IASC Date: Sat, 23 Feb 2013 13:58:23 +0000 Thanks, Barbara. I will process your request. Patrick From: Barbara Dluzewska Sent: Saturday, February 23, 2013 6:22 AM To: Miller, Patrick Cc: Lech Mankiewicz Subject: Re: Asteroid Discovery -- IASC Hi, Patrick, We would like to name our asteroid Paweł Maksym , on the memory of the young astronomer that died a few days ago. Lech will write to you about him. We hope, everything will go well, with best regards, Barbara W dniu 4/16/2012 6:28 AM, Miller, Patrick pisze: Hi, Barbara. Welcome to the IASC Discovery Hall of Fame!! You and your students have an asteroid discovery that has been numbered and placed into the world's official minor bodies catalog. On April 6, 2007, you and your students M. Bogowicz, P. Jasinski, D. Swierczewska, & B. Dluzewska from Czacki High School discovered 2007 GA4, which is now numbered as 315166. You are the third school from Poland with this accomplishment.

[28] H. Povenmire, *Graze Observers Handbook*, JSB Enterprises, Indian Harbour Beach, FL, (1975).
[29] http://www.youtube.com/watch?v=Gutyrl6GDN4&list=UUAhmVcvUFXjQoGgliyo2G9g&index=9

[30] http://www.youtube.com/watch?v=m1Al4bZUrHw&list=UUAhmVcvUFXjQoGgliyo2G9g&index=10

[31] H. Araki et al., Science **323**, 897 (2009).

**Total lunar occultations**

This is the starting activity of occultation astronomy, already mentioned in the Almagest of Ptolemy and used to discover the secular acceleration of the Moon.[32]

These observations are nowadays very important to help in the calibration of the Kaguya profile of the limb of the Moon, in order to be used in the Baily's beads method of the measurement of the solar diameter.[33]

Paweł Maksym published the occultation of the Pleiades, M45, of July 18, 2009 observed from Andrespol, the birth city of his wife Katarzyna.

This kind of occultation allows a rapid test of the Kaguya profile, as implemented, for example, in the Occult 4 software, because the stars disappear and reappear at different position angles.[34]

The final type of lunar occultation studied by Paweł Maksym is the occultation of a planet, namely Saturn, on March 2 and May 22, 2007.[35]

The analysis of the light curve of gaseous planets and their satellites gives interesting information, which can be fruitfully exploited especially in educational contexts as at the Planetarium of Łódź, in which Paweł worked up to 2010, and the Pope Silvester II Observatory of Bukowiec.

**Meteor videos**

The observation of the Orionids meteor shower was made with the camera of the observing station n. 29 at Łódź Planetarium (PAV29) of the Polish Fireball Network.[36]

The first fireball presented in the video was recorded on October 22, 2007 at 4:00:52 UT and it lasted about one second, showing two consecutive luminosity peaks.

Probably inspired by these observations, Paweł Maksym made two videos of the movement of clouds over the city of Łódź, using a VX21000 Sony camcorder with a polarising filter and

---

[32] F. R. Stephenson, *Historical Eclipses and Earth's Rotation*, Cambridge University Press, Cambridge, UK, (1997).

[33] C. Sigismondi, Science in China Series G: Physics, Mechanics and Astronomy, **52**, 1773 (2009).

[34] C. Sigismondi, Journal of the Korean Physical Society, **56**, 1694 (2010).

[35] http://www.youtube.com/watch?v=M3KQdbqgpRw&list=UUAhmVcvUFXjQoGgliyo2G9g&index=18

http://www.youtube.com/watch?v=v3nk6wvnrCA&list=UUAhmVcvUFXjQoGgliyo2G9g&index=17

[36] M. Wiśniewski, et al., EPSC Abstracts Vol. 7 EPSC2012-497 (2012).
http://meetingorganizer.copernicus.org/EPSC2012/EPSC2012-497.pdf

Raynox 0,5x wide angle lenses, with excellent artistic results and noteworthy scientific and didactic interest.[37]

**Solar eclipses**

Paweł Maksym observed the eclipse of March 29, 2006 in Egypt and presented his results to the XXV ESOP meeting held in Leiden (NL), under the affiliation of the Ary Sternfeld Planetarium and Astronomical Observatory in Łódź and the Polish Association of Amateur Astronomers Department of Position and Occultations.[38]

**The Pope Silvester II Observatory of Bukowiec**

This has been his major realisation, with the largest impact on society: an observatory founded and built with his great capability of creating interests, motivations and strengths.

Located in the surroundings of Łódź,[39] it was erroneously considered a competitor of the planetarium of Łódź, but this article is written also for demonstrating the contrary, as I made indirectly in an interview with Paweł in November 2011 when he visited the Vatican and he was my guest.[40]

The inauguration took place on May 21, 2010 with a large gathering of personalities and local people. Paweł wrote an enthusiastic paper in the first issue of the Journal for Occultation Astronomy,[41] and put 252 photos of the event onto the web.[42] He honoured me with an invitation onto the scientific committee[43] and to take part in the inauguration. The city council carefully considered the proposal to name the Observatory (which is located in the grounds of the Nicolas Copernicus school) after Silvester II, the Pope-astronomer of the year 1000;[35] they accepted it with enthusiasm.

---

[37] http://www.youtube.com/watch?v=iCw5vupXe7g&list=UUAhmVcvUFXjQoGgliyo2G9g  (18 March 2008, with sunset)

http://www.youtube.com/watch?v=c_HWc6Xoda0&list=UUAhmVcvUFXjQoGgliyo2G9g (26 March 2008)

[38] http://www.doa-site.nl/esop25/papers.html#Anchor-Pawe-9166

[39] https://maps.google.com/maps/ms?ie=UTF8&hl=pl&source=embed&t=h&msa=0&msid=204447440296630627728.000485b74f186d3be2e16&ll=51.726603,19.637375&spn=0.246258,0.727158&z=11

[40] http://www.youtube.com/watch?v=_kfz57773-o&list=UUAhmVcvUFXjQoGgliyo2G9g&index=4

[41] P. Maksym, JOA **1**, 11 (2011), available on http://www.iota-es.de/JOA/joa2011_1.pdf

[42] https://plus.google.com/photos/118080826487159559391/albums/5474542097173125041?banner=pwa

[43] See on www.oabukowiec.pl   the page linked to the English info includes the list of the Honour Scientific Committee.

The ceremony was accompanied by a *pièce de théatre* that he wrote for the occasion. The text of this theatrical work in the Polish language, with an English abstract, has been published in the first issue of Gerbertus, the electronic academic journal dedicated to Silvester II studies and to medieval science and astronomy.[44] Gerbert of Aurillac-Silvester II, Galileo, Copernicus and Boghdan Paczyński are the subjects of this play, and the historical portrayal helped the public to be acquainted with this forgotten scientist, Silvester II, and with his astronomical skills.

The Pope Silvester II Observatory is active mainly in lunar, asteroidal and TNO occultations, and observations made there were already included in some Minor Planet Circulars.

Name: Pope Silvester II's Astronomical Observatory in Bukowiec

Nearest city: Łódź, Poland

Latitude: N 51° 41' 28.23"

Longitude: E 19° 40' 32.80" Altitude: 214.1 m

Remarks (GPS, map): WGS 84, GPS, Altitude (MSL) obtained with geodesic map

website http://www.oabukowiec.pl/

in Minor Planet Center format

Longitude: 19.675778

Cos: 0.619899

Sin: 0.784681

Polish Name: Obserwatorium Astronomiczne im. Papieza Sylwestra II w Bukowcu

**Editorial activity**

Another important activity that Paweł Maksym assisted in was the editorial board of the Journal for Occultation Astronomy. Since the resurrection of the dormant Occultation Newsletter was discussed and agreed at the XXIX ESOP meeting in York, Paweł took a key role in collating and publishing the European contributions to this journal. The international scientific community acknowledged his leading role in this way, and his cultural and scientific level. In the short time he had at his disposal he made an excellent job of work, contributing to the very high level of the new journal of IOTA members. His death leaves a big hole that will be very difficult to fill, especially in the years to come, when the majority of current IOTA members will reach retirement age.

---

[44] P. Maksym, Gerbertus **1** 198 (2010), http://www.icra.it/gerbertus/2010/Gerbertus_1-pp198-212-Maksym.pdf

The material in the Polish language that Paweł Maksym has left on the web is much more detailed than these guidelines about his research activity, and many other works are recommended in order to gather all his contributions and present them to a wider public.

As an example I have found the conference Obserwacje zjawisk zakryciowych, The observation of occultation phenomena,[45] where the main occultations of the last decade are discussed, namely the grazing occultation of HIP 9369 and Jupiter's atmosphere of October 10, 1999, the occultation of the star HIP 19388 by (345) Tercidina and the solar eclipse observed by Paweł et al in Egypt.

Indeed, the colleagues of Paweł Maksym, who were acquainted with him and the Polish language can find more material either from private sources or on the web. They could, hopefully, publish his thesis on lunar occultations and the introductory book on astronomy for children that he was writing in the last weeks of his life, whilst he was preparing for the surgery of February 4th 2013.

It is recommended to do it also in the Polish language, because he cared so much about spreading this culture in his country. It is possible to repeat also the title and an abstract in English, so this text can be indexed to be reached by the international search engines and the international science community. The material could be published in the Journal for Occultation Astronomy but also on the arxiv.org website, where only the structure with title, abstract, text and references are required, and with an opportune second abstract and title in English that can be easily accessed worldwide.

On the occasion of the International Year of Astronomy Paweł Maksym wrote a tutorial in Polish on the use of the "Galileoscope".[46] Such an instrument was intended to celebrate the 400 years of the first observations with the telescope and Paweł presented its potential in the observation of the sky.

**Perspectives**

The seminal works of Paweł Maksym in the field of occultation astronomy and for his country have been acknowledged in the international contexts of ESOP meetings where Paweł participated since the XIX Symposium held in Łódź in August 2000 at the age of 17. He was present at all ESOP meetings, and he was the organiser of the 2009 Symposium in Niepolomice, near Krakow. In the last ESOP meeting XXXI in Pescara[47] he presented a talk about the new

---

[45] http://urania.pta.edu.pl/pliki/kruszwica/zakrycia_ogolna.pdf

[46] http://www.slideshare.net/JacekKupras/galileoskop-przewodnik-obserwacyjny

[47] www.icranet.org/clavius2012

GPS time inserter from Poland, showing the progress of the basic task of an IOTA observer, the accurate timing of observations.

Both amateur and professional astronomers need this kind of applied research, because the issue of time keeping is of paramount importance in occultation astronomy, and the issue of the association of a time interval with a video frame requires the maximum attention and reliability.

It would be nice if sessions at the next ESOP meetings would welcome the papers of IOTA historians with contributions to outline and remember the great work of Paweł.

The publication of his miscellanea and his profiles as seen from his former collaborators are also suggested. This volume would be one of the grounds on which the newborn Silvester II Observatory and the newborn Karol Josef Maksym, who arrived the day before the partial solar eclipse of January 4th 2011, can continue their path to the future, with a more clear identity of their scientific and genetic father.

**Acknowledgements:** To Alexander Pratt for his kind help in the correction of the text and its development and to Felipe Braga-Ribas for the useful discussions on TNOs, and in particular on the last Varuna's occultation.

**References of Paweł Maksym's works on the web**

<http://www.youtube.com/watch?v=JEs4LRH7tAo> TV news about his death February 13, 2013

<http://www.youtube.com/watch?v=M3KQdbqgpRw&list=UUAhmVcvUFXjQoGgliyo2G9g&index=18> Saturn occultation of March 2, 2007

<http://www.youtube.com/watch?v=v3nk6wvnrCA&list=UUAhmVcvUFXjQoGgliyo2G9g&index=17> Saturn occultation of May 22, 2007

<http://www.youtube.com/watch?v=ol_3aWgIyK4&list=UUAhmVcvUFXjQoGgliyo2G9g> Orionid meteor shower

<http://www.youtube.com/watch?v=m1Al4bZUrHw&list=UUAhmVcvUFXjQoGgliyo2G9g&index=10> Zakrycie brzegowe/Graze of ZC 1298 - 6.4 mag - 13.04.2008

<http://www.youtube.com/watch?v=Gutyrl6GDN4&list=UUAhmVcvUFXjQoGgliyo2G9g&index=9>  Zakrycie brzegowe/Graze of SAO77818 - 21.04.2007 mv=6.7  SONY HC 96E

<http://www.youtube.com/watch?v=A-Mf7V1KdLQ&list=UUAhmVcvUFXjQoGgliyo2G9g&index=7>

M45 total lunar occultation xz76103 xz76128 Merope xz76135  xz76145 175 198 (in neg. view)

http://www.youtube.com/watch?v=yCxDoEj9Ktw&list=UUAhmVcvUFXjQoGgliyo2G9g

Watec Camera 902H specifications and test TYC 0727-01424-1 mag 11.4 by 1144 Oda 8" newton EQ5 on 4 set 2008 (a negative asteroidal occultation)

http://www.youtube.com/watch?v=wh7lerr0pMM&list=UUAhmVcvUFXjQoGgliyo2G9g&index=6 Occultation of star TYC 2934-00106-1, 9.4 mag by asteroid (71) Niobe on Monday, 24 August 2009 in Bukowiec

http://www.youtube.com/watch?v=ZrsAGEw6ukc&list=UUAhmVcvUFXjQoGgliyo2G9g&index=5 Occultation of HIP 48340 (8.2 Mag) by asteroid 173 Ino on June 7, 2011 at the Silvester II Observatory in Bukowiec

http://www.youtube.com/watch?v=iCw5vupXe7g&list=UUAhmVcvUFXjQoGgliyo2G9g
Clouds over Łódź, with sunset and "eclipse" by clouds

http://www.youtube.com/watch?v=_kfz57773-o&list=UUAhmVcvUFXjQoGgliyo2G9g&index=4
Interview made by Paweł Maksym with C. Sigismondi

Contributions on Minor Planet Circulars:

http://www.minorplanetcenter.net/iau/ECS/MPCArchive/2007/MPC_20070926.pdf

http://www.minorplanetcenter.net/iau/ECS/MPCArchive/2008/MPC_20080122.pdf

http://www.minorplanetcenter.net/iau/ECS/MPCArchive/2009/MPC_20091231.pdf

http://www.minorplanetcenter.net/iau/ECS/MPCArchive/2011/MPC_20110715.pdf

http://www.minorplanetcenter.net/iau/ECS/MPCArchive/2011/MPC_20110912.pdf

P. Maksym, JOA **1**, 11 (2011), available on http://www.iota-es.de/JOA/joa2011_1.pdf

P. Maksym, Gerbertus **1** 198 (2010), available on http://www.icra.it/gerbertus/2010/Gerbertus_1-pp198-212-Maksym.pdf

http://urania.pta.edu.pl/pliki/kruszwica/zakrycia_ogolna.pdf    The observation of occultation phenomena